\documentclass[twocolumn]{aastex63}
\usepackage{CJK}
\usepackage{amsmath}
\usepackage[normalem]{ulem}

\received{}
\revised{}
\accepted{}
\submitjournal{ApJ}

\shorttitle{oblique tearing instability}
\shortauthors{Shi et al.}
\graphicspath{{./}{figures/}}

\begin{document}
\begin{CJK*}{UTF8}{gbsn}
\title{Oblique tearing mode instability: guide field and Hall effect}

\correspondingauthor{Chen Shi}
\email{cshi1993@ucla.edu}

\author[0000-0002-2582-7085]{Chen Shi (时辰)}
\affiliation{Earth, Planetary, and Space Sciences \\
University of California, Los Angeles \\
Los Angeles, CA 90095, USA}

\author[0000-0002-2381-3106]{Marco Velli}
\affiliation{Earth, Planetary, and Space Sciences \\
University of California, Los Angeles \\
Los Angeles, CA 90095, USA}

\author{Fulvia Pucci}
\affiliation{Laboratory for Atmospheric and Space Physics \\
University of Colorado, Boulder\\
Boulder, CO 80303, USA}

\author[0000-0003-2880-6084]{Anna Tenerani}
\affiliation{Department of Physics\\
The University of Texas at Austin\\
Austin, TX 78712, USA}

\author[0000-0002-5782-0013]{Maria Elena Innocenti}
\affiliation{Department of Mathematics, Centre for mathematical Plasma Astrophysics,\\
University of Leuven (KULeuven), \\
Celestijnenlaan 200B, Leuven, Belgium}



\begin{abstract}

The tearing mode instability is one important mechanism that may explain the triggering of fast magnetic reconnection in astrophysical plasmas such as the solar corona and the Earth's magnetosphere. In this paper, the linear stability analysis of the tearing mode is carried out for a current sheet in the presence of a guide field, including the Hall effect. 
We show that the presence of a strong guide field does not modify the most unstable mode in the two-dimensional wave vector space orthogonal to the current gradient direction, which remains the fastest growing parallel mode. With the Hall effect, the inclusion of a guide field turns the non-dispersive propagation along the guide field direction to a dispersive one. The oblique modes have a wave-like structure along the normal direction of the current sheet and a strong guide field suppresses this structure while making the eigen-functions asymmetric.

\end{abstract}



\section{Introduction}\label{sec:introduction}
Magnetic reconnection is a process that allows the topology of the magnetic field to change in a plasma, leading to the conversion of magnetic energy into thermal and kinetic energy. It is thought to be the cause of various explosive phenomena in astrophysical plasmas such as coronal mass ejections (CMEs) and geomagnetic storms.

Since the original Sweet-Parker model of reconnection \citep{Sweet1958,Parker1957} an open question has remained, namely how the release of magnetic energy can proceed as fast as seen observationally. In the Sweet-Parker model, the reconnection rate inside a resistive current sheet scales as $R \sim S_L^{-1/2}$, where $R = V_{in}/V_{out}$ measures the speed of the plasma inflow carrying the magnetic flux into the reconnection region, $S_L = LV_{A}/\eta$ is the Lundquist number, where $L$ is the scale length of the current sheet, $\eta$ is the magnetic diffusivity, $V_A$ is the upstream Alfv\'en speed, and $V_{out} \sim V_A$. As pointed out by \citet{Parker1957}, in most astrophysical plasmas, $S_L$ is extremely large (e.g. $S_L > 10^8$ in the solar atmosphere), meaning that the reconnection rate is too slow to explain explosive phenomena in such astrophysical plasmas.

In the last two decades, great progress was achieved in not only the fast kinetic-scale reconnection \citep[e.g.][]{birn2001geospace}, but also understanding the triggering of fast reconnection through the tearing mode instability first analyzed by \citet{Furth1963}. The tearing mode instability 
inside an infinite (1D) current sheet has a maximum growth rate $\gamma \tau_{a} \sim S_a^{-1/2} $ where $S_a=aV_A/\eta$ is the Lundquist number measured by the thickness of the current sheet and $\tau_a = a/V_A$ is the characteristic Alfv\'en time. Although it seems from the above relation that the growth rate of tearing instability is very low when $S_a$ is large, it was noticed that, in a two-dimensional current sheet, i.e. a current sheet with finite aspect ratio $a/L$, the re-normalized growth rate has a different scaling relation with the Lundquist number $S_L$: $\gamma \tau_L \sim S_L^{\alpha}$ where $\tau_L = L/V_A$ and $\alpha$ depends on the aspect ratio of the current sheet \citep{TajimaandShibata2002,Loureiroetal2007}. Especially, for a Sweet-Parker type current sheet whose aspect ratio obeys $a/L \sim S_L^{-1/2}$, one obtains $\gamma \tau_L \sim S_L^{1/4}$. This positive scaling relation leads to enormous growth rates for large Lundquist number, meaning that a thinning current sheet will break up due to the fast-growing tearing instability before it ever reaches the Sweet-Parker aspect ratio. \citet{PucciandVelli2013} argued that once a scaling aspect ratio $a/L\sim S_{L}^{-1/3}$ is reached and the growth rate of the most unstable mode becomes independent of $S_L$, any further current sheet thinning will be disrupted by reconnection.  They called this limit ``ideal tearing'' (IT).  \citet{Teneranietal2015} and \citet{Landietal15} confirmed this scenario by means of resistive-MHD simulations showing that in a collapsing current sheet, fast plasmoid-generation occurs when the aspect ratio of the current sheet reaches the IT threshold. In addition, the subsequent evolution leads to a nonlinear recursive reconnection stage \citep[see also][]{ShibataandTanuma2001}. More recently, \citet{Shietal2018} showed
how the decreasing Lundquist number of the higher-order current sheets generated during the recursive X-point collapse between islands quenches
the regeneration, while \citet{Shietal2019, Papinietal19} discussed the role of he Hall, or ion kinetic effects, in increasing the X-point separatrix angle, accelerating reconnection while quenching subsequent plasmoid formation.
 
\begin{figure}[htb!]
    \centering
    \includegraphics[scale=0.6]{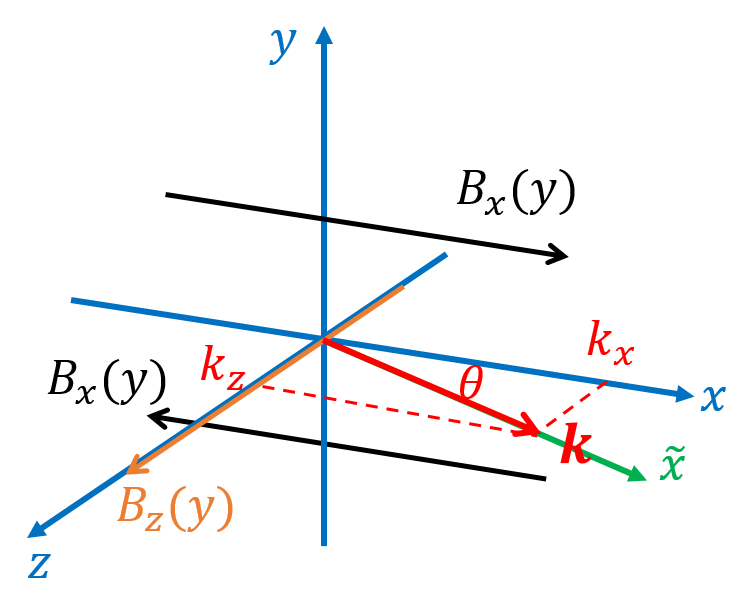}
    \caption{Illustration of the coordinate system and background magnetic field.}
    \label{fig:coordinates}
\end{figure}

From the linear point of view, other progress made in the last several years includes the study of the oblique tearing mode in the case of a strong guide field and the introduction of kinetic effects in the IT scenario.  On the former \citet{Baalrudetal2012} showed that, in the so-called constant-$\psi$ regime (where $\psi$ refers to the magnetic flux function), corresponding to relatively large wave-numbers along the unperturbed tearing unstable field component ($B_x (y) $), the fastest growing modes have finite $k_z$, where $k_z$ is the wave number along the guide field direction as illustrated in Figure \ref{fig:coordinates}. For simplicity we will hereafter refer to $x$ as the parallel direction and $k_x$ as the parallel wave number.

Concerning the Hall effect, \citet{Puccietal2017} extended the ideal tearing theory to include its effect on a planar sheet, and calculated the modified critical aspect ratio which triggers the ideal tearing mode.
In this study, we carry out a linear analysis of the tearing mode instability in a more general configuration. We allow a guide field with arbitrary strength and include the Hall effect. We numerically solve the linear eigenvalue problem for the oblique tearing modes. We show that, although a guide field results in a resonant surface departing from the parallel direction in the constant-$\psi$ regime, the overall fastest growing mode in the $(k_x, k_z)$ plane is still parallel, i.e. it remains the same one as in the case without guide field. With the Hall effect, the guide field generates a dispersive $\omega(k_z)$ where $\omega$ is the oscillation frequency. The paper is organized as follows. In Section \ref{sec:linear_equation}, we present the linear equation set that we solve. In Section \ref{sec:results} we show the numerical solutions of the linear equation set. In Section \ref{sec:conclusion}, we conclude this study and discuss possible future works.

\section{Linear MHD equation set for oblique tearing mode}\label{sec:linear_equation}
We start from the three-dimensional MHD equation set with Hall term and resistive term:
\begin{subequations}
    \begin{equation}
        \frac{\partial \rho}{\partial t} + \nabla \cdot \left( \rho \mathbf{U} \right) = 0
    \end{equation}
    \begin{equation}
        \rho \left( \frac{\partial \mathbf{U}}{\partial t} + \mathbf{U} \cdot \nabla \mathbf{U} \right) = - \nabla P + \mathbf{J \times \mathbf{B}}
    \end{equation}
    \begin{equation}
        \frac{\partial \mathbf{B}}{\partial t} = \nabla \times \left( \mathbf{U} \times \mathbf{B} \right) +  \eta \nabla^2 \mathbf{B} - \nabla \times \left( \frac{\mathbf{J} \times \mathbf{B}}{ne} \right)
    \end{equation}
\end{subequations}
where $(\rho,\mathbf{U},\mathbf{B},P)$ are the density, velocity, magnetic field, and scalar pressure respectively, $\mathbf{J}= \nabla \times \mathbf{B}/\mu_0$ is the current density and $n$ is the number density. The background flow is assumed to be 0 everywhere ($\mathbf{U_0}\equiv 0$) and the background density is assumed to be uniform $\rho_0 \equiv 1$. The background magnetic field is of the form
\begin{equation}
    \mathbf{B_0} = B_x(y) \hat{e}_x + B_z(y) \hat{e}_z
\end{equation}
where $B_z$ is the guide field and the background pressure $P_0 = P(y)$ such that 
\begin{equation}
    P(y) + \frac{1}{2} \left[ B_x^2(y) + B_z^2(y)\right] = Const
\end{equation}
The above configuration is a solution to the Hall-MHD equation set with zero resistivity. For finite resistivity ($\eta >0$), the magnetic field diffuses which modifies the growth rate of the instability. But for very small resistivity considered in this paper, the diffusion rate of the background field is low so we can neglect it. In this study, we will restrict the background magnetic field to be a Harris current sheet \citep{Harris1962} plus a uniform guide field
\begin{equation}\label{eq:background_Bfield}
    B_x(y) = B_0 \tanh \left( \frac{y}{a} \right) ,  B_z(y) = B_g
\end{equation}
with $B_0 \equiv 1 $, $a \equiv 1$, and $B_g$ being a varying parameter. This configuration implies that the Lorentz force is balanced by the pressure gradient force induced by the inhomogeneous temperature, which is observed in the Earth's magnetotail \citep[e.g.][]{lu2019effects,lu2019turbulence}. In applications to the Sun, e.g. the solar flare problem, the current sheet is possibly force-free:
\begin{equation}\label{eq:background_Bfield_force_free}
\begin{aligned}
    &B_x(y) = B_0 \tanh \left( \frac{y}{a} \right) ,\\
    &B_z(y) = \sqrt{B_g^2 + B_0^2 \mathrm{sech}^2\left( \frac{y}{a} \right)}
\end{aligned}
\end{equation}
but we do not expect the two magnetic field profiles (Eq (\ref{eq:background_Bfield}) \& (\ref{eq:background_Bfield_force_free})) will lead to significantly different growth rates of the tearing instability.

We write perturbations in the form:
\begin{equation}
    \mathbf{u} = \mathbf{u}(y) e^{\gamma t + i \mathbf{k} \cdot \mathbf{x}}, \quad \mathbf{b} = \mathbf{b}(y) e^{\gamma t + i \mathbf{k} \cdot \mathbf{x}}
\end{equation}
where $\mathbf{k} = k_x \hat{e}_x + k_z \hat{e}_z$. We assume incompressibility 
\begin{equation}
    \nabla \cdot \mathbf{u} = 0
\end{equation}
for simplicity since it was shown by \citet{Furth1963} that compressibility has negligible effect on the tearing mode. The normalized equation set for $\mathbf{u}$ and $\mathbf{b}$ is written as (taking curl of the 1st-order momentum equation to get rid of pressure $p_1^T$)
\begin{subequations}
    \begin{equation}
        \gamma \nabla \times \mathbf{u} = \nabla \times \left( \mathbf{B} \cdot \nabla \mathbf{b} + \mathbf{b} \cdot \nabla \mathbf{B} \right)
    \end{equation}
    \begin{equation}
        \gamma \mathbf{b} = \mathbf{B} \cdot \nabla \mathbf{u} - \mathbf{u} \cdot \nabla \mathbf{B} + \frac{1}{S} \nabla^2 \mathbf{b} - \gamma d_i \nabla \times \mathbf{u}
    \end{equation}
\end{subequations}
Here $S = a V_A / \eta$ is the Lundquist number and $d_i=\frac{c}{a}\sqrt{\frac{\varepsilon_0 m_p}{n e^2}}$ is the normalized ion inertial length (ion skin depth) where $m_p$ is the proton mass, $c$ is the speed of light, $e$ is the elementary charge, and $\varepsilon_0$ is the vacuum electric permittivity. We adopt the method by \citet{CaoandKan1991} to simplify the equation, i.e. we rotate the coordinate system with respect to the $y$-axis such that the new $\tilde{x}$ axis is aligned with the wave vector:
\begin{equation}
    \mathbf{k} = k \hat{e}_{\tilde{x}}
\end{equation}
where $k=\sqrt{k_x^2 + k_z^2}$. Then the problem becomes essentially 2D because $\partial_{\tilde{z}} \equiv 0$. Figure \ref{fig:coordinates} illustrates the coordinate system as well as the background magnetic field. In the new coordinate system we get the new form of the background magnetic field
\begin{equation}
    \begin{aligned}
    \tilde{B}_{\tilde{x}}(y) & = B_x(y) \cos \theta + B_z(y) \sin \theta, \\
    \tilde{B}_{\tilde{z}}(y) & = - B_x(y) \sin \theta + B_z(y) \cos \theta
    \end{aligned}
\end{equation}
where $\theta = \arctan \left( k_z/ k_x \right)$. The closed equation set for $(u_{y},b_y,b_{\tilde{z}})$ is
\begin{subequations}\label{eq:linear_eq_set}
    \begin{equation}\label{eq:linear_eq_uy}
        \gamma \left( u_y^{\prime\prime} - k^2 u_y  \right) = k \left[ \tilde{B}_{\tilde{x}} b_y^{\prime\prime} - \left(\tilde{B}_{\tilde{x}}^{\prime\prime} + k^2 \tilde{B}_{\tilde{x}} \right)b_y \right]
    \end{equation}
    \begin{equation}\label{eq:linear_eq_by}
        \begin{aligned}
            \frac{1}{S} \left( b_y^{\prime \prime} - k^2 b_y \right) = & \gamma b_y + k \tilde{B}_{\tilde{x}} u_y \\
             &- i d_i k \left( -k \tilde{B}_{\tilde{x}} b_{\tilde{z}} + \tilde{B}_{\tilde{z}}^\prime b_y \right)
        \end{aligned}
    \end{equation}
    \begin{equation}\label{eq:linear_eq_bz}
        \begin{aligned}
            \frac{1}{S} \left( b_{\tilde{z}}^{\prime\prime} - k^2 b_{\tilde{z}} \right) = & \left( \gamma + \frac{k^2 \tilde{B}_{\tilde{x}}^2}{\gamma} \right) b_{\tilde{z}} - \frac{k \tilde{B}_{\tilde{x}} \tilde{B}_{\tilde{z}}^\prime}{\gamma } b_y  \\
            & + \tilde{B}_{\tilde{z}}^\prime u_y - i d_i \frac{\gamma}{k} \left( u_y^{\prime\prime} - k^2 u_y  \right) 
        \end{aligned}
    \end{equation}
\end{subequations}
where we have replaced $ib_y$ with $b_y$ and prime means the derivative in the $y$ direction. $u_{\tilde{x}}$ and $b_{\tilde{x}}$ can be derived from the divergence-free conditions of $\mathbf{u}$ and $\mathbf{b}$ and the equation for $u_{\tilde{z}}$ is 
\begin{equation}
    \gamma u_{\tilde{z}} = i \left( k \tilde{B}_{\tilde{x}} b_{\tilde{z}} - \tilde{B}_{\tilde{z}}^\prime b_y \right)
\end{equation}
Note that Eq (\ref{eq:linear_eq_set}) is general, i.e. we can arbitrarily choose functions $B_x(y)$ and $B_z(y)$ such as Eq (\ref{eq:background_Bfield_force_free}) but in this study we use Eq (\ref{eq:background_Bfield}).

It is immediately seen that in the case $d_i=0$, Eq (\ref{eq:linear_eq_set}) is purely real, i.e. there are no propagating modes since the solution of $\gamma$ is real. In addition, the $d_i=0$ condition decouples Eq (\ref{eq:linear_eq_bz}) from the other two equations so the eigenvalue $\gamma$ can be fully determined by Eq (\ref{eq:linear_eq_uy}) \& (\ref{eq:linear_eq_by}). In this case the background magnetic field appears only in the form: $k \tilde{B}_{\tilde{x}} = \mathbf{k} \cdot \mathbf{B_0} $. If $d_i > 0$, in general $\gamma$ is complex, meaning that the modes are propagating. But there is a special case $\tilde{B}_{\tilde{z}}^\prime = 0$ when $b_{\tilde{z}}$ has an exactly $\pi/2$ phase-difference with $u_y$ and $b_y$ and thus by doing the transformation $ib_{\tilde{z}} \rightarrow b_{\tilde{z}}$, Eq (\ref{eq:linear_eq_set}) becomes purely real and so does $\gamma$. This is the case when a mode is parallel ($k_z = 0$) and the guide field is uniform, e.g. the case considered by \citet{Puccietal2017}. In reality, when $k_z=0$, a uniform $B_z$ has no effect on Eq (\ref{eq:linear_eq_set}) as only $B_z^\prime$ enters the equation.

Last, we need to specify the boundary condition in order to solve the eigenvalue problem. Far from the center of the current sheet, we have all the derivatives of $\tilde{B}_{\tilde{x}}$ and $\tilde{B}_{\tilde{z}}$ in Eq (\ref{eq:linear_eq_set}) to be 0 and it is easy to find that the solutions decay exponentially with distance as $\exp \left(- k \left| y\right| \right)$. This is the same boundary condition as the classic 2D tearing mode.

\section{Results}\label{sec:results}

\begin{figure*}[htb!]
    \centering
    \includegraphics[scale=0.55]{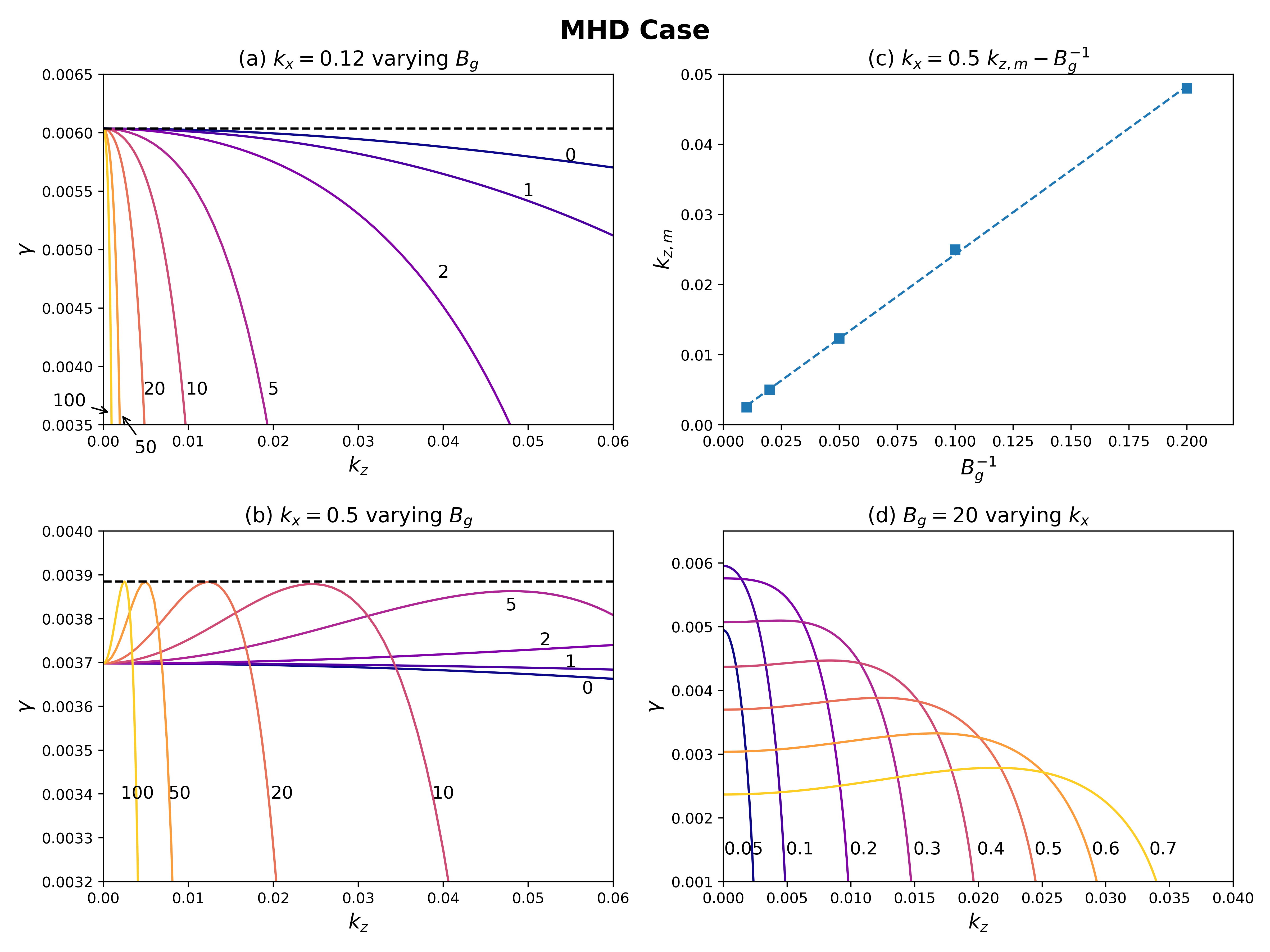}
    \caption{Results for $S=10^4$ and $d_i=0$. Panel (a) \& (b): $\gamma-k_z$ curves at $k_x=0.12$ and at $k_x=0.5$. Colors represent guide field strength which is annotated in the plots. Panel (c): $k_z$ of the fastest growing mode at $k_x=0.5$ as a function of $B_g^{-1}$. Panel (d): $\gamma-k_z$ curves at different $k_x$ (annotated in the plot) with guide field strength $B_g=20$.}
    \label{fig:gamma_kz_varying_Bg_di_0}
\end{figure*} 

We use the boundary-value-problem (BVP) solver implemented in the Python library SciPy \citep{Virtanenetal2020Scipy} to solve Eq (\ref{eq:linear_eq_set}). The solver adopts a 4th order collocation algorithm with the control of residuals \citep[ref.][]{Kierzenkaetal2001BVP,Ascheretal1994numerical} and is able to solve the eigenvalue and eigen-functions simultaneously. Unlike previous works, e.g. \citet{Baalrudetal2012}, which use $(k, \theta)$ to denote the wave vector, we present our results in $(k_x, k_z)$ space. In this study, we fix $S=10^4$, a value large enough for astrophysical applications (corresponding to $S_L=10^8$ for a Sweet-Parker current sheet) and not too large so that it is not very expensive to solve Eq (\ref{eq:linear_eq_set}). The domain used for solving the equation is $y \in [-15,15]$.

\subsection{MHD case}\label{sec:results_MHD}
We first consider the MHD case, i.e. $d_i=0$, for which the problem is purely real. In Figure \ref{fig:gamma_kz_varying_Bg_di_0}, we show the dispersion relation $\gamma- k_z$ at two fixed $k_x$: $k_x=0.12$ in Panel (a) and $k_x=0.5$ in Panel (b). $k_x=0.12$ corresponds approximately to the fastest-growing parallel mode. 
In each of the two panels, different curves represent different guide field strength $B_g$. Dark to light colors correspond to small (0) to large (100) $B_g$ as written in the plots. From Panel (a) ($k_x=0.12$, fastest-growing parallel mode), we observe that $\gamma$ in general declines as $k_z$ increases and increasing $B_g$ speeds up the decline of $\gamma$ with $k_z$. The fastest growing mode is always the parallel one ($k_z = 0$). On the contrary, Panel (b) ($k_x=0.5$, non-fastest-growing parallel mode) shows very different results from Panel (a). For small $B_g$ ($B_g \lesssim 1$), $\gamma-k_z$ is monotonically decreasing. As $B_g$ increases ($B_g \gtrsim 2$), the $\gamma-k_z$ curve transits from monotonic to concave and the fastest growing mode is no longer the parallel one but instead located at the new resonant surface. In Panel (c) we show $k_z$ of the fastest growing mode ($k_{z,m}$) at $k_x=0.5$ (peaks of curves in Panel (b)) as a function of $1/B_g$. It can be observed that $k_{z,m}$ is proportional to $B_g^{-1}$, consistent with the prediction of the resonant surface:
\begin{equation}
    \mathbf{k} \cdot \mathbf{B_0} = 0
\end{equation}
which gives
\begin{equation}
    k_z = -\frac{k_x B_x}{B_g}
\end{equation}

\begin{figure}[htb!]
    \centering
    \includegraphics[scale=0.5]{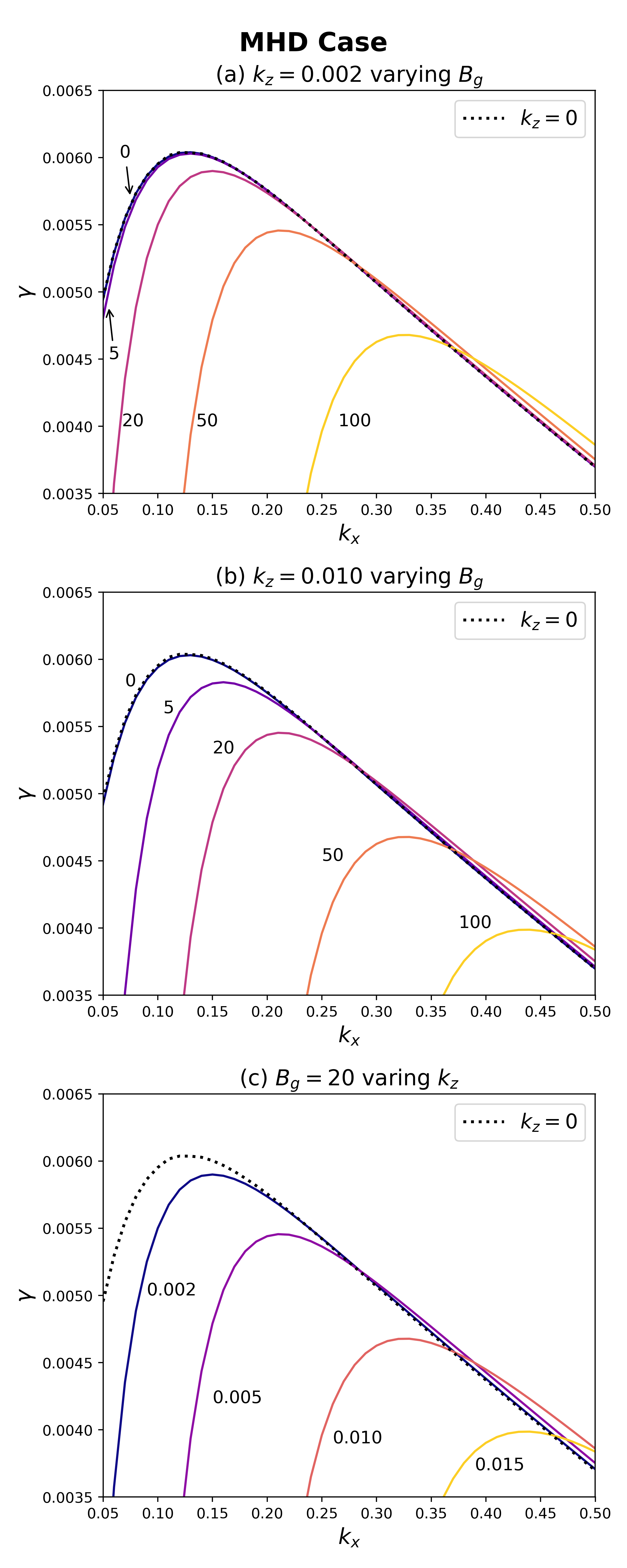}
    \caption{Results for $S=10^4$ and $d_i=0$. Panel (a) \& (b): $\gamma-k_x$ curves at $k_z=0.002$ and at $k_z=0.01$. Colors represent different $B_g$ as annotated near the curves. Panel (c): $\gamma-k_x$ curves for $B_g=20$ and varying $k_z$, which is annotated near the curves. The dashed curve in each panel is $k_z = 0$, i.e. parallel modes, for reference.}
    \label{fig:gamma_kx_varing_Bg_kz=combined_di=0}
\end{figure}

We stress that, although a strong guide field leads to an increase in $\max \left(\gamma(k_z) \right)$ at fixed $k_x$ in the constant-$\psi$ regime, the increase is small. As can be seen from Panel (b) of Figure \ref{fig:gamma_kz_varying_Bg_di_0}, $\max \left(\gamma(k_z)\right)$ increases from about $3.7 \times 10^{-3}$ to about $3.88 \times 10^{-3}$, i.e. only by $\sim 5 \%$ as $B_g$ goes from $B_g=0$ to $B_g=100$. Furthermore, the increase of $\max \left( \gamma (k_z) \right)$ only occurs in the constant-$\psi$ regime (large $k_x$), but not at the most unstable $k_x$, as shown in Panel (a) of Figure \ref{fig:gamma_kz_varying_Bg_di_0}. In Panel (d) we plot $\gamma-k_z$ curves for $B_g=20$ at different $k_x$.  The positive slope of the $\gamma-k_z$ curves at large $k_x$ due to the strong guide field does not compensate for the overall decrease in the values of $\gamma$ for such values of $k_x$ . Thus, a strong guide field $B_g$ cannot change the fastest growing mode in the $(k_x,k_z)$ plane: it is always the most unstable parallel mode. Here, it is helpful to clarify again that by ``parallel'' we mean parallel to $\hat{e}_x$, i.e. the anti-parallel magnetic field direction, though in the limit $B_g \rightarrow \infty$ the true parallel direction becomes the guide field direction. To support the above conclusion, in Figure \ref{fig:gamma_kx_varing_Bg_kz=combined_di=0}, we plot $\gamma-k_x$ curves at $k_z=0.002$ in Panel (a) and at $k_z=0.01$ in Panel (b). Colors represent different $B_g$ and the black dashed curve is $k_z=0$, i.e. parallel modes, for reference. Note that a uniform $B_g$ has no influence on the parallel modes. 
At a fixed $k_z$, although a strong guide field rises the $\gamma-k_x$ curve slightly in the constant-$\psi$ regime, it lowers the curve significantly at smaller $k_x$. With increasing $B_g$, the fastest growing $k_x$ is shifted toward the right, i.e toward larger values, while the peak growth rate declines rapidly. In Panel (c) of Figure \ref{fig:gamma_kx_varing_Bg_kz=combined_di=0}, we plot $\gamma-k_x$ curves for a fixed $B_g=20$ but varying $k_z$. It can be seen by comparing Panels (b) \& (c) that increasing $k_z$ with constant $B_g$ has nearly identical effect as increasing $B_g$ with constant $k_z$. This is because that in Eq (\ref{eq:linear_eq_set}), in the case of a uniform guide field, all terms containing $B_g$ are of the form $k_z B_g$. From this plot, we can see that, the maximum growth rate $\max \left( \gamma(k_x) \right)$ as a function of $k_z$ is monotonically decreasing, supporting the conclusion that the fastest-growing mode in the 2D $(k_x,k_z)$ plane is always the most unstable parallel mode even with a strong guide field $B_g$. Unless the current sheet is very short along $x$, such that $k_x$ is limited to the constant-$\psi$ regime \citep{Leakeetal2020,VelliandHood1989,Vellietal1990}, or the system size along $z$ is finite \citep{huang2009effects}, we do not expect the most unstable mode to be oblique, though this does not imply that oblique modes do not become fundamental in the nonlinear evolution (see, e.g. \citet{Landietal08}).

\subsection{Hall-MHD case}
\begin{figure*}[htb!]
    \centering
    \includegraphics[scale=0.6]{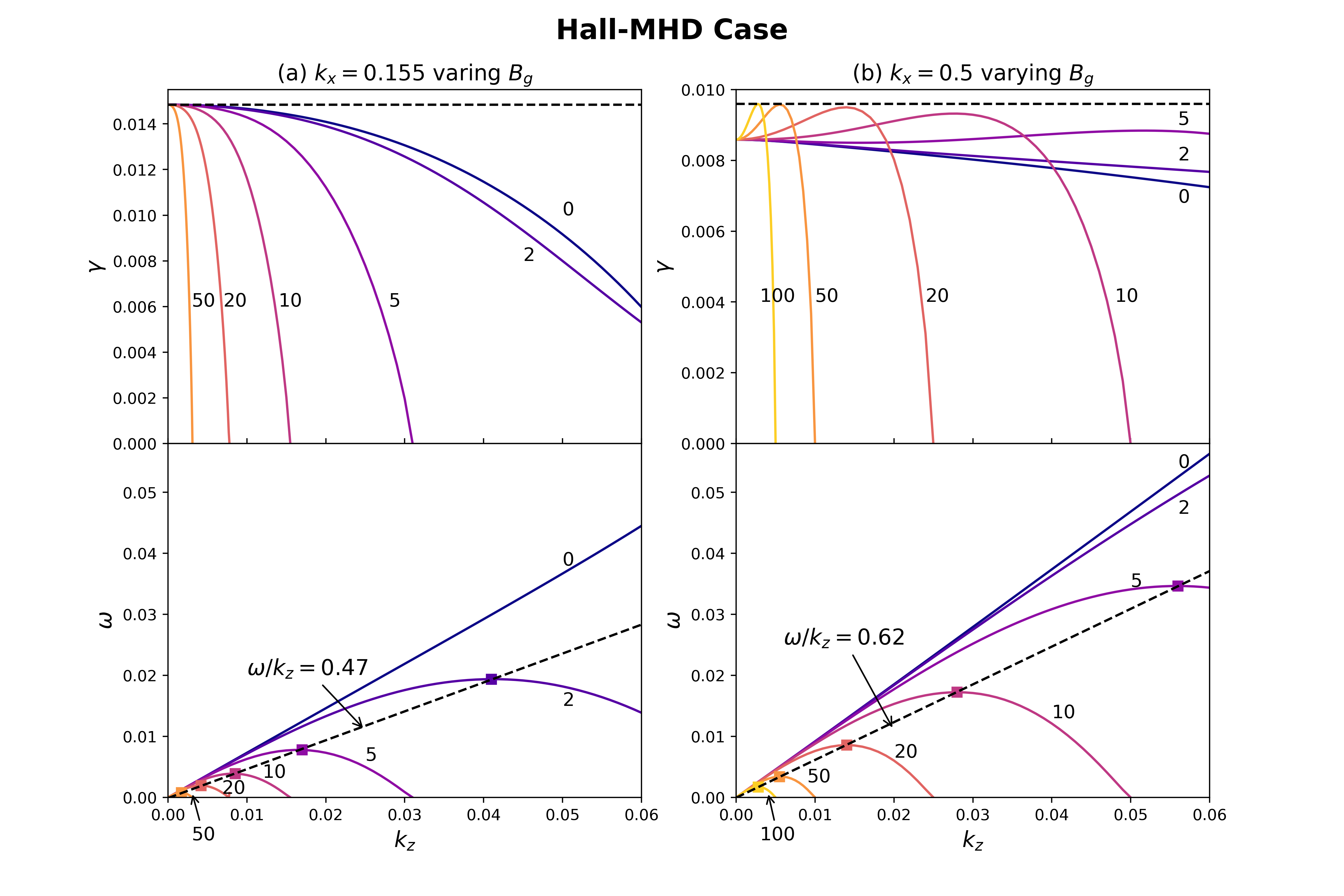}
    \caption{$\gamma - k_z$ (top) and $\omega - k_z$ (bottom) curves for $S=10^4$, $d_i=1.0$. Here $\gamma$ and $\omega$ are the real and imaginary parts of the complex $\gamma$ in Eq (\ref{eq:linear_eq_set}). Column (a) is $k_x=0.155$, roughly the most unstable parallel mode and Column (b) is $k_x=0.5$. Colors represent guide field strength $B_g$ as annotated in the plots. Squares in the bottom panels are the peaks of the $\omega-k_z$ curves and dashed lines are linear-fits of the squares.}
    \label{fig:gamma_omega_kz_varying_Bg_di=1}
\end{figure*}

We then consider the case with finite ion inertial length. In this case, $\gamma$ in Eq (\ref{eq:linear_eq_set}) is complex and we decompose it into real and imaginary parts $\gamma - i \omega$ where $\gamma$ is the growth rate and $\omega$ is the oscillation frequency. Indeed, in the solar corona, the ion inertial length is much smaller than the size of the typical macroscopic current sheet. However, in the recursive reconnection scenario \citep[e.g.][]{Teneranietal2015,Landietal15}, the high-order current sheets may approach the ion inertial length \citep[e.g.][]{Shietal2019}. In other environments, e.g. the Earth's magnetotail and magnetopause, the thickness of the current sheet is usually on the same order of the ion inertial length. Thus, it is necessary to explore how the Hall effect modifies the tearing mode.

\begin{figure}[htb!]
    \centering
    \includegraphics[scale=0.6]{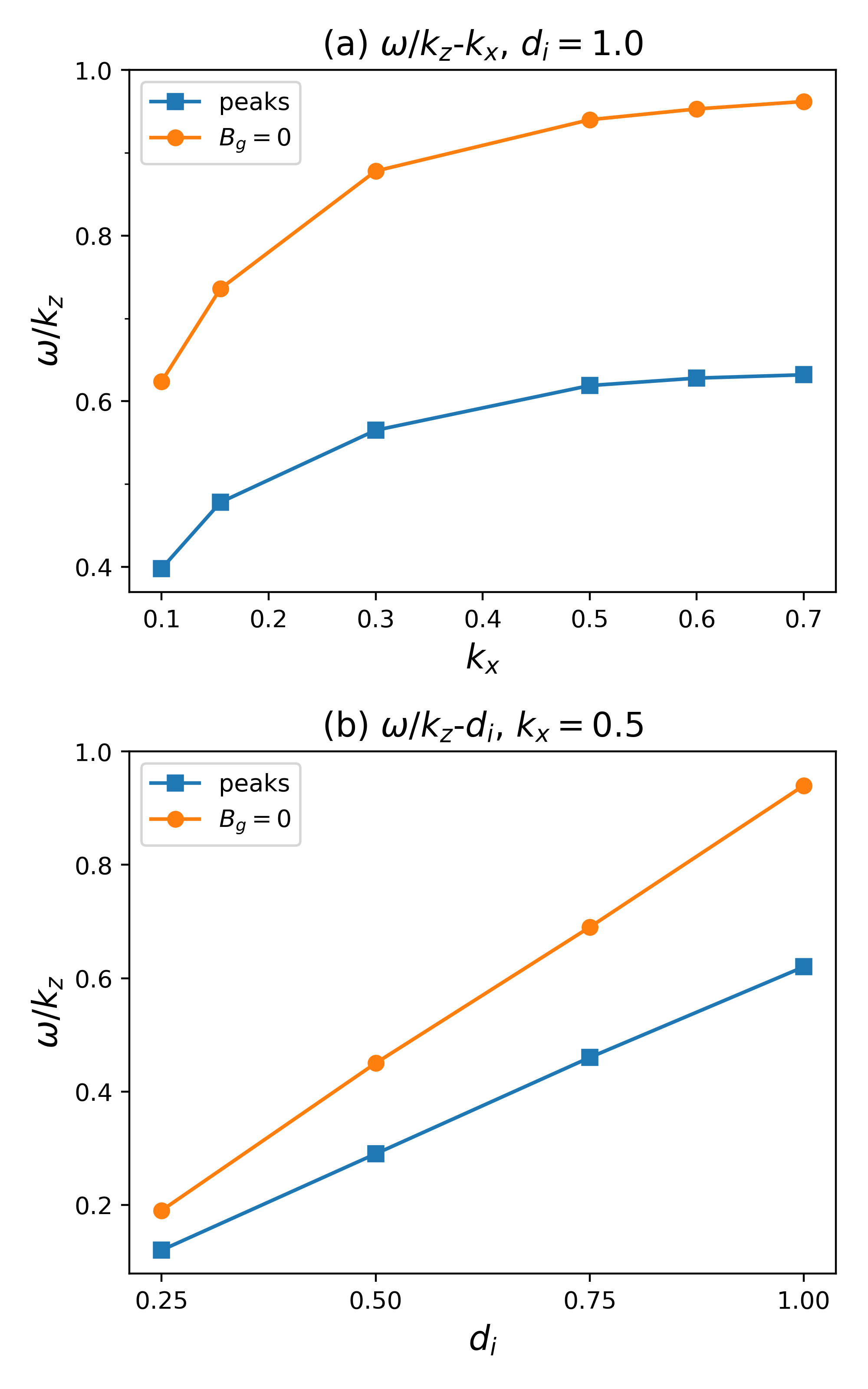}
    \caption{Slopes of the lines threading the peaks of the $\omega-k_z$ curves (e.g. black dashed lines in Figure \ref{fig:gamma_omega_kz_varying_Bg_di=1}) (blue) and slopes of the $\omega-k_z$ curves with $B_g=0$ (orange) as functions of $k_x$ with $d_i=1.0$ (a) and functions of $d_i$ with $k_x=0.5$ (b). The Lundquist number is $S=10^4$.}
    \label{fig:slope_omega_kz}
\end{figure}


In Figure \ref{fig:gamma_omega_kz_varying_Bg_di=1} we plot $\gamma-k_z$ (top) and $\omega-k_z$ (bottom) curves for $d_i = 1$ and different guide field strength at $k_x = 0.155$ (a), which corresponds roughly to the most unstable parallel mode, and $k_x=0.5$ (b). The behavior of $\gamma$ is similar to the MHD case. For small $k_x$, $B_g$ does not rise the $\gamma-k_z$ curve. For large $k_x$, a slight rise of the $\gamma-k_z$ is observed around the resonant surface $k_z \propto B_g^{-1}$ and the increase of $\max \left( \gamma (k_z) \right)$ has an asymptotic value as $B_g$ increases. Thus, the conclusion made in the MHD case is not modified. Note that, by comparing the top panel on Column (b) of Figure \ref{fig:gamma_omega_kz_varying_Bg_di=1} with Panel (b) of Figure \ref{fig:gamma_kz_varying_Bg_di_0}, we can see that the growth rate is larger with a finite $d_i$, as already reported by \citep{Puccietal2017}. $\omega(k_z)$ has an interesting behavior: For weak guide field, $\omega$ is almost a linear function of $k_z$. 
Especially, for $B_g=0$, $\omega-k_z$ is exactly a straight line, i.e. the modes are non-dispersive along $z$ direction. As $B_g$ increases, $\omega(k_z)$ is no longer monotonic but shows a decline with $k_z$ after reaching a peak value. In bottom panels of Figure \ref{fig:gamma_omega_kz_varying_Bg_di=1}, we mark the peak of each individual curve by a square and the black dashed line in each panel is the linear fit of the squares. The extrapolation of each dashed line goes through the origin and the squares are aligned on the dashed line, indicating that the maximum $\omega$, i.e. the mode with fastest phase speed along $x$ as $k_x$ is fixed, has a phase speed along $z$ which is independent of $B_g$ when $B_g$ is large. As can be seen from the figure, the slope of the dashed line (written in the plots) changes with $k_x$. In Panel (a) of Figure \ref{fig:slope_omega_kz} we present this slope $(\omega/k_z)_m$ as a function of $k_x$ in blue squares, for $S=10^4$ and $d_i=1$. It is seen that $(\omega/k_z)_m$ increases with $k_x$ and reaches an asymptotic value (0.63 in this case). 
For reference we also plot the slope $\omega/k_z$ for $B_g=0$ in orange circles and we can see that the two slopes are highly correlated. Note that, the $k_z$ corresponding to $\max \left( \omega (k_z) \right)$ does not necessarily correspond to $\max \left( \gamma (k_z) \right)$. In reality, for small $k_x$, $\omega-k_z$ curves show peaks while $\gamma-k_z$ curves are monotonically decreasing. 

\begin{figure*}[htb!]
    \centering
    \includegraphics[scale=0.45]{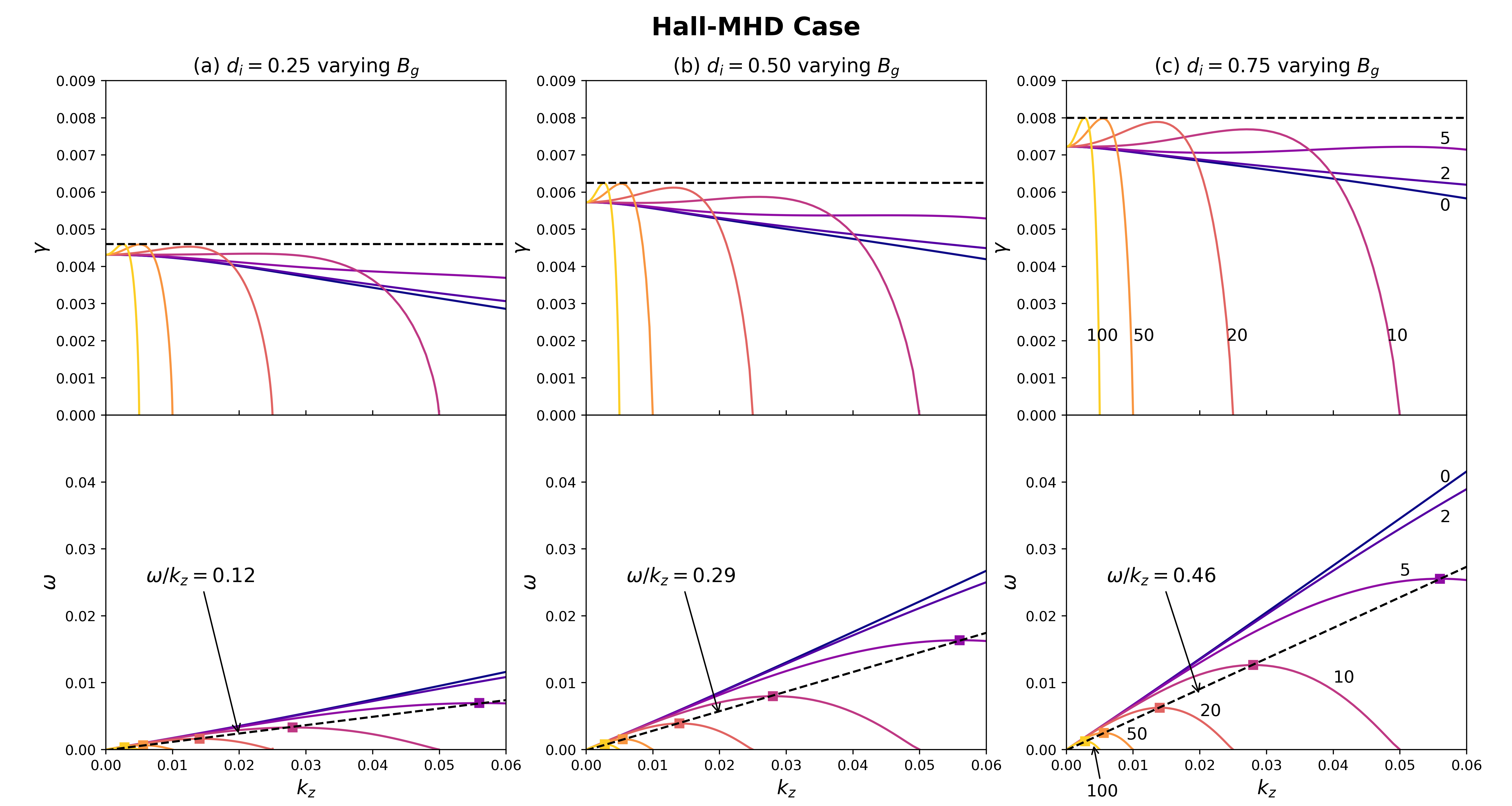}
    \caption{$\gamma-k_z$ (top) and $\omega - k_z$ (bottom) curves with varying $B_g$, for $S=10^4$, $k_x=0.5$ and $d_i=0.25$ (a), $d_i=0.50$ (b) and $d_i=0.75$ (c). The values of $B_g$ are annotated in Column (c) near the curves but not in Column (a) \& (b) as the layout of the curves in Column (a) \& (b) is similar to Column (c).}
    \label{fig:gamma_omega_kz_varying_Bg_different_di}
\end{figure*}

In Figure \ref{fig:gamma_omega_kz_varying_Bg_different_di} we show $\gamma-k_z$ (top) and $\omega-k_z$ (bottom) curves with varying guide field strength at $k_x=0.5$ for $d_i=0.25$ (a), $d_i=0.50$ (b), and $d_i=0.75$ (c). We annotate the corresponding $B_g$ values near the curves in Column (c). In Column (a) \& (b), the layout of the curves is very similar to Column (c) so we do not annotate the $B_g$ values separately.
In general, the growth rate increases with $d_i$ as expected and the $\omega-k_z$ curves are also rised by $d_i$. However, the cut-off $k_z$, i.e. $k_z$ at which $\gamma$ and $\omega$ drop to zero, do not change with $d_i$. In Panel (b) of Figure \ref{fig:slope_omega_kz}, we plot the slope $(\omega/k_z)_m$ as a function of $d_i$ for $k_x=0.5$ in blue squares and we see the $(\omega/k_z)_m - d_i$ relation is linear. Similar to Panel (a), we plot the slope $\omega/k_z$ for $B_g=0$ in orange circles and obviously its relation with $d_i$ is also linear. The behavior of $\omega$ can be partially understood by inspecting Eq (\ref{eq:linear_eq_by}) where the balance between the following two terms
\begin{equation}
    \gamma b_y \sim i d_i k \tilde{B}^\prime_{\tilde{z}} b_y
\end{equation}
gives the estimate of $\omega$:
\begin{equation}\label{eq:estimate_omega}
    \omega \sim k d_i \tilde{B}^\prime_{\tilde{z}} = k_z d_i  B_x^\prime.
\end{equation}
The above relation indicates that $\omega/k_z$ is proportional to $d_i$ and is not significantly affected by the guide field strength. It also explains why $\omega(k_z)$ is non-dispersive for $B_g=0$. As will be discussed in Section \ref{sec:conclusion}, Eq (\ref{eq:estimate_omega}) actually reflects the motion of ions that serve as current carriers along the guide field direction. If we take into account the first term inside the bracket on the r.h.s. of Eq (\ref{eq:linear_eq_by}), the above estimate is modified:
\begin{equation}
    \omega \sim d_i \left[ - k^2 \tilde{B}_{\tilde{x}} \frac{b_{\tilde{z}}}{b_y} + k_z B_x^\prime \right]
\end{equation}
which may explain the nonlinear $\omega(k_z)$ relation with finite $B_g$ and the nonlinear $k_x$-dependence of $(\omega/k_z)_m$ as shown in Panel (a) of Figure \ref{fig:slope_omega_kz}.

\begin{figure*}[htb!]
    \centering
    \includegraphics[scale=0.7]{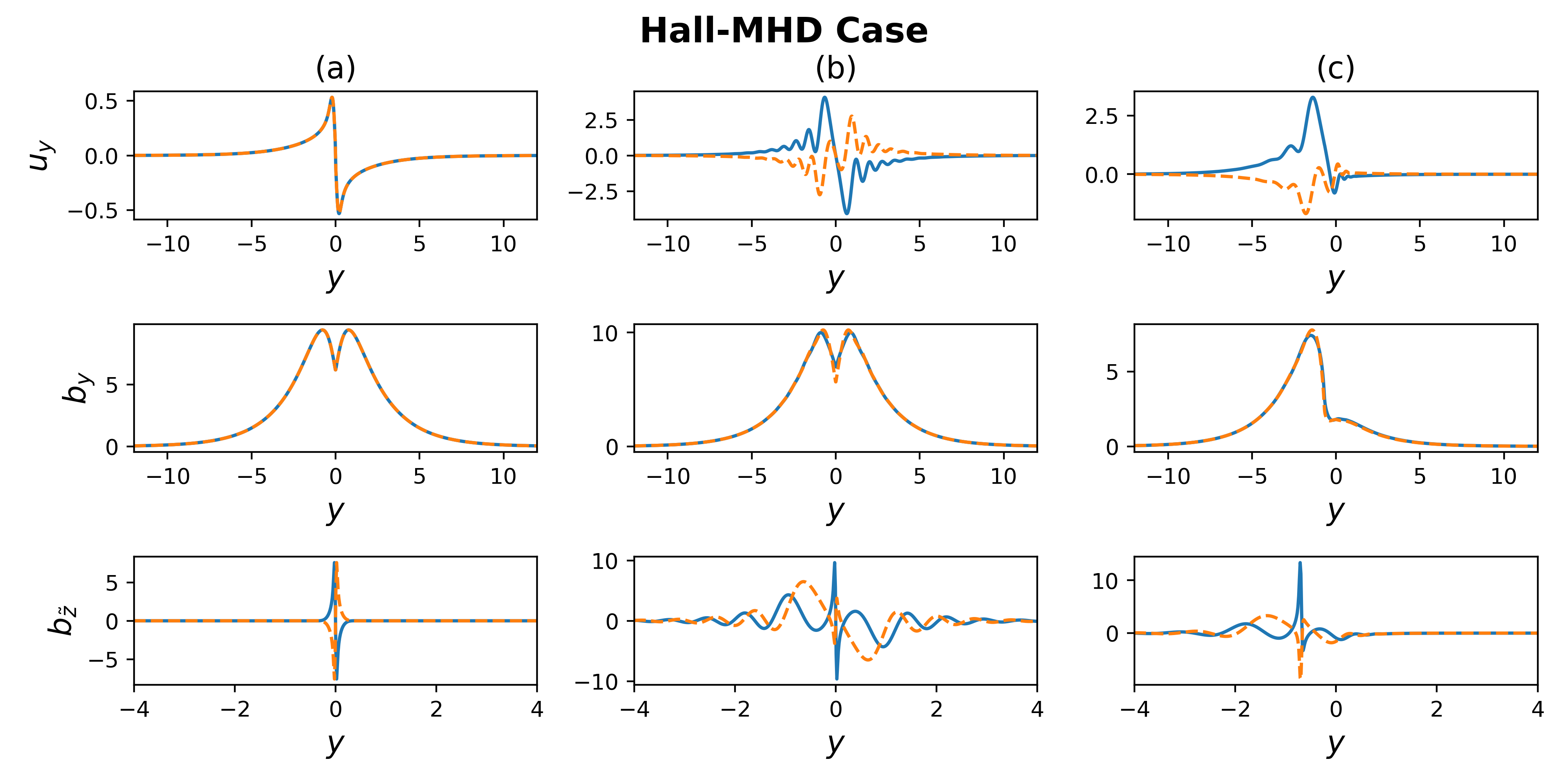}
    \caption{Solution of eigen-functions $u_y$ (top row), $b_y$ (middle row) and $b_{\tilde{z}}$ (bottom row) for  $S=10^4$, $d_i=1.0$ and $k_x=0.5$. Blue and orange curves are real and imaginary parts of each eigen-function. Left column (a) is for $k_z=0$ and arbitrary $B_g$, middle column (b) is for $k_z=0.06$ and $B_g=0$, right column (c) is for $k_z=0.06$ and $B_g=5$.}
    \label{fig:eigen_function}
\end{figure*}

In Figure \ref{fig:eigen_function}, we show the eigen-functions solved from Eq (\ref{eq:linear_eq_set}) for $S=10^4$, $d_i=1.0$ and $k_x=0.5$ (corresponding to Column (b) of Figure \ref{fig:gamma_omega_kz_varying_Bg_di=1}). 
Top row shows $u_y$, middle row shows $b_y$ and bottom row shows $b_{\tilde{z}}$. Note that the range of abscissa is smaller for $b_{\tilde{z}}$ because the inner layer of $b_{\tilde{z}}$ is much thinner than those of the other two quantities \citep{Puccietal2017}. Blue and orange curves in each panel are real and imaginary parts respectively. Column (a) is for $k_z=0$, i.e. parallel mode thus $B_g$ can be any value, as discussed in Section \ref{sec:linear_equation}. 
Column (b) is for $k_z=0.06$ and $B_g = 0$, i.e. oblique mode without guide field. Column (c) is for $k_z=0.06$ and $B_g = 5$, i.e. oblique mode with guide field. 
The main point of Figure \ref{fig:eigen_function} is that, with Hall effect, the eigen-functions of the oblique mode show strong oscillation along $y$, as can be seen from Column (b). The oscillation is caused by the term $  b_{y}^{\prime\prime}/S \sim  -id_ik \tilde{B}^\prime_{\tilde{z}} b_{y}$ in Eq (\ref{eq:linear_eq_by}) from which we can estimate a wave number along $y$ to be
\begin{equation}
    k_y \sim (1-i) \sqrt{\frac{1}{2} S k_z d_i B_{x}^\prime} 
\end{equation}
in the sense that $b_y \sim \exp \left(k_y y \right)$. With a guide field, as shown in Column (c), the eigen-functions become asymmetric in $y$ as expected \citep{Baalrudetal2012}. In addition, by comparing Column (b) and (c), we see that the strong guide field suppresses the $y$-oscillation, through the first term inside the bracket on the r.h.s. of Eq (\ref{eq:linear_eq_by}). The phenomenon of $y$-oscillation was not reported by the previous study on oblique tearing mode with finite ion inertial length \citep{CaoandKan1991}. The reason was unknown but it might be that \citet{CaoandKan1991} carried out linear simulations to solve the problem and the resolution (not stated in \citep{CaoandKan1991}) was not enough. Recently, \citet{Akccayetal2016} carried out two-fluid simulations of the oblique tearing mode and in their simulations this oscillation was seen (see their Figure 4).

\section{Discussion and Conclusion}\label{sec:conclusion}
In this study, we carried out linear stability calculation of the oblique tearing mode with both guide field and Hall effect. We derived the generally-applicable linear equation set for incompressible tearing mode instability (Eq (\ref{eq:linear_eq_set})). We show that, although a guide field leads to a non-parallel resonant surface $k_z \propto 1/B_g$ in the constant-$\psi$ regime (Panel (c) of Figure \ref{fig:gamma_kz_varying_Bg_di_0}), the most unstable mode in the $(k_x,k_z)$ space is not changed: it is still the fastest-growing parallel mode. The increase in $\max \left( \gamma(k_z) \right)$ at a fixed $k_x$ due to the guide field is limited to a small fraction (Panel (b) of Figure \ref{fig:gamma_kz_varying_Bg_di_0}). The $\max \left( \gamma(k_x) \right)$ is a monotonically decreasing function of $k_z$, i.e. increasing the wave number along the guide field direction always lowers the largest growth rate of the tearing mode (Panel (c) of Figure \ref{fig:gamma_kx_varing_Bg_kz=combined_di=0}). The presence of Hall effect, i.e. ion kinetic effect, does not modify the above conclusion regarding the growth rate of tearing mode although a finite ion inertial length increases the growth rate in general. The existence of the Hall effect makes the oblique tearing mode propagate rather than purely grow. Without a guide field, the $\omega-k_z$ relation is non-dispersive, i.e. all the oblique modes at a fixed $k_x$ propagate along the guide field at the same speed. A strong guide field turns the linear $\omega-k_z$ line into a non-monotonic curve and the peak of the $\omega(k_z)$ curve has a $\omega/k_z$ value independent of $B_g$ (Figure \ref{fig:gamma_omega_kz_varying_Bg_di=1} \& \ref{fig:gamma_omega_kz_varying_Bg_different_di}). That is to say, the fastest $x$-propagating mode has a phase speed along the guide field independent of the guide field strength but depends only on $k_x$ and $d_i$ (Figure \ref{fig:slope_omega_kz}). Last, with the Hall effect, the oblique tearing mode has a propagating component cross the current sheet ($y$ direction) and this component is suppressed by a strong guide field (Figure \ref{fig:eigen_function}).

Our main result is that, even with a strong guide field, the most unstable tearing mode is still parallel. However, in astrophysical context like the solar atmosphere, the condition for this result to be applicable is not necessarily satisfied. As we have mentioned in Section \ref{sec:results_MHD}, when the length of the current sheet is finite, e.g. several current sheet thicknesses, such that $k_x$ is confined to large values, the tearing mode only exists in the constant-$\psi$ regime. This is the so called ``line-tying'' scenario, which is a good model to describe the magnetic field of the coronal loop anchored deeply in the photosphere. If the guide field, i.e. the field threading the magnetic loops, is strong, the most unstable mode may be oblique instead of parallel. In this case, we expect that the growth of tearing mode generates a series of plasma ``patches'' along the guide field direction rather than flux tubes. Our results also reveal that a strong guide field significantly confines the range of $k_z$ (to small values) in which the tearing mode can grow. This implies that, with a strong guide field, the system size along the guide field direction needs to be large enough, e.g. $L_z \approx O(10^2) a$ for $S=10^4$ and $B_g=5$ (Panel (a) of Figure \ref{fig:gamma_kz_varying_Bg_di_0}), to include growing tearing modes. Thus, in case that the guide field is also line-tied, e.g. the progenitor of CME \citep{chen2011coronal}, the tearing mode may be stabilized by a strong guide field. In simulations of the coronal loop \citep[e.g.][]{rappazzo2010shear}, it was observed that only after the initially-uniform guide field becomes highly-sheared due to the transverse motion of the photospheric plasma that tearing instability is triggered. This can also be explained by our result that with a strong guide field and finite size along the guide field, the tearing instability is suppressed.

For reconnection happening in the Earth's magnetosphere, the current sheet thickness is on the ion scale. At the magnetopause, a guide field is usually present and it is observed that the X-line of the reconnection site is of finite size and spreads along the guide field direction \citep[e.g.][]{zou2018spreading}. The spreading of the X-line is explained by either the motion of the current carrier, either electrons or ions, or the propagation of Alfv\'en wave \citep{shepherd2012guide}. The first mechanism gives an estimate of the spreading speed (see Eq (1) of \citep{shepherd2012guide}):
\begin{equation}
    v \sim \frac{1}{ne} \mathbf{J} \sim \frac{B_{rec}}{ne\mu_0 a} \sim \frac{d_i}{a} V_{A}
\end{equation}
where $V_A = B_{rec}/\sqrt{\mu_0  n m_i}$ is the upstream Alfv\'en speed defined by the reconnecting component of the magnetic field. This is consistent with our estimate Eq (\ref{eq:estimate_omega}) (note that our equation adopts normalized quantities):
\begin{equation}
    \frac{\omega}{k_z} \sim d_i B_x^\prime 
\end{equation}
Our result reveals that, in linear stage, the propagation speed of the tearing mode along the out-of-plane direction is on the order of Alfv\'en speed based on the anti-parallel component of the background magnetic field (refer to Figure \ref{fig:gamma_omega_kz_varying_Bg_di=1}-\ref{fig:gamma_omega_kz_varying_Bg_different_di}) and is highly dependent on $d_i$, or alternatively the current sheet thickness. As the current sheet thickness decreases toward the ion inertial length, the propagation of the mode speeds up. In general, with a strong guide field, the out-of-plane propagation speed is lower compared with the case without guide field and is weakly-dependent on the guide field strength. This contrasts the second scenario of \citet{shepherd2012guide} that the spreading speed is the Alfv\'en wave speed along the guide field:
\begin{equation}
    v = V_{A,z} = \frac{B_{g}}{\sqrt{\mu_0 n m_i}}
\end{equation}
which predicts a spreading speed proportional to the guide field strength. In reality, \citet{zou2018spreading}, by analyzing conjugate spacecraft and ground radar data, found that with a strong guide field, the spreading speed of the magnetopause X-line can only be explained by the current-carrier mechanism and is smaller than the Alfv\'en speed along the guide field. The failure of acquiring such a propagation speed from our linear calculation implies the necessity to carry out nonlinear 3D Hall-MHD simulations in the future.

\acknowledgments
This research was funded in part by the FIELDS experiment on the Parker Solar Probe spacecraft,
designed and developed under NASA contract NNN06AA01C and the NASA
Parker Solar Probe Observatory Scientist grant NNX15AF34G. It was also supported by the NSF-DOE Partnership in Basic Plasma Science and Engineering award n. 1619611. M.E.I.'s work is supported by an FWO (Fonds voor Wetenschappelijk Onderzoek – Vlaanderen) postdoctoral fellowship.

%


\software{SciPy \citep{Virtanenetal2020Scipy},
    Matplotlib \citep{Hunter2007Matplotlib}
          }






\bibliography{references}{}
\bibliographystyle{aasjournal}


\end{CJK*}
\end{document}